\documentclass[fleqn,usenatbib]{mnras}

\usepackage{newtxtext,newtxmath}

\usepackage[T1]{fontenc}

\DeclareRobustCommand{\VAN}[3]{#2}
\let\VANthebibliography\thebibliography
\def\thebibliography{\DeclareRobustCommand{\VAN}[3]{##3}\VANthebibliography}

\usepackage{graphicx}	
\usepackage{amsmath}	

\usepackage{xspace}

\newcommand{\jwst}{\textit{JWST}\xspace}
\newcommand{\ha}{\mathrm{H}\,\alpha}

\newcommand{\starburstlong}{MGTDR1\,J100034.18$+$015733.8\xspace}
\newcommand{\sfglong}{MGTDR1\,J100000.46$+$022256.2\xspace}
\newcommand{\agnlong}{MGTDR1\,J095940.32$+$022331.5\xspace}

\newcommand{\starburstshort}{MGT\,J10003$+$01573\xspace}
\newcommand{\sfgshort}{MGT\,J10000$+$02225\xspace}
\newcommand{\agnshort}{MGT\,J09594$+$02233\xspace}

\title[Three radio-selected SFGs at $z=4.9$--$5.6$]{MIGHTEE/COSMOS-3D: The discovery of three spectroscopically confirmed radio-selected star-forming galaxies at $\mathbf{z=4.9}$--5.6}

\author[R. G. Varadaraj et al.]{R. G. Varadaraj,$^{1}$\thanks{E-mail: rohan.varadaraj@physics.ox.ac.uk}
A. Saxena,$^{1,2}$
S. Fakiolas,$^{1}$
I. H. Whittam,$^{1,3}$
M. J. Jarvis,$^{1,3}$
R. A. Meyer,$^{4}$
\newauthor
C. L. Hale,$^{1,5}$
K. Kakiichi,$^{6,7}$
M. Li,$^{8}$
J. B. Champagne,$^{9}$
B. Jin$^{10}$,
Z. J. Li,$^{11,12}$
M. Shuntov$^{6,7}$
\\
$^{1}$Astrophysics, Department of Physics, University of Oxford, Keble Road, Oxford, OX1 3RH, UK\\
$^{2}$Department of Physics and Astronomy, University College London, Gower Street, London WC1E 6BT, UK\\
$^{3}$Department of Physics and Astronomy, University of the Western Cape, Robert Sobukwe Road, 7535 Bellville, Cape Town, South Africa\\
$^{4}$Department of Astronomy, University of Geneva, Chemin Pegasi 51, 1290 Versoix, Switzerland\\
$^{5}$Institute for Astronomy, University of Edinburgh, Royal Observatory Edinburgh, Blackford Hill, Edinburgh, EH9 3HJ, UK\\
$^{6}$ Cosmic Dawn Center (DAWN), Denmark\\
$^{7}$ Niels Bohr Institute, University of Copenhagen, Jagtvej 128, 2200 Copenhagen, Denmark\\
$^{8}$Department of Astronomy, Tsinghua University, Beijing 100084, China\\
$^{9}$Steward Observatory, University of Arizona,
933 N Cherry Ave, Tucson, AZ 85721, USA\\
$^{10}$Department of Astronomy, University of Michigan, 1085 S. University Ave., Ann Arbor, MI 48109, USA\\
$^{11}$Chinese Academy of Sciences South America Center for Astronomy (CASSACA),
National Astronomical Observatories of China (NAOC),
CAS, 20A Datun\\ Road, Beijing 100012, China\\
$^{12}$School of Astronomy and Space Sciences, University of Chinese Academy of Sciences, Beijing 100049, China\\
}

\date{Accepted XXX. Received YYY; in original form ZZZ}

\pubyear{\the\year{}}

\begin{document}
\label{firstpage}
\pagerange{\pageref{firstpage}--\pageref{lastpage}}
\maketitle

\begin{abstract}
Radio observations offer a dust-independent probe of star formation and active galactic nucleus (AGN) activity, but sufficiently deep data are required to access the crossover luminosity between these processes at high redshift ($z>4.5$).
We present three spectroscopically confirmed high-redshift radio sources (HzRSs) detected at 1.3 GHz at $z=4.9$--$5.6$, with radio luminosities spanning $L_{\rm 1.3 \, GHz}\approx2$--$5\times10^{24} \, \rm W \, Hz^{-1}$.
These sources were first identified as high-redshift candidates through spectral energy distribution (SED) fitting of archival \textit{Hubble}, \jwst NIRCam+MIRI, and ground-based photometry, and then spectroscopically confirmed via the $\ha$ emission line using wide-field slitless spectroscopy from \jwst COSMOS-3D.
The star formation rates (SFRs) measured from SED fitting, the $\ha$ flux, and the 1.3 GHz luminosity, span $\sim100$--$1800\, M_{\odot} \, \rm yr^{-1}$, demonstrating broad agreement between these SFR tracers. We find that these three sources lie either on or $0.5$--$1.0$\,dex above the star-forming main sequence at $z=4$--$6$ and have undergone a recent burst of star formation.
The sources have extended rest-UV/optical morphologies with no evidence for a dominant point source component, indicating that an AGN is unlikely to dominate their rest-UV and optical emission.
Two of the sources have complex, multi-component rest-frame UV/optical morphologies, suggesting that their starbursts may be triggered by merging activity.
These HzRSs open up a new window towards probing radio emission powered by star formation alone at $z> 4.5$, representing a remarkable opportunity to begin tracing star formation, independent of dust, in the early Universe.
\end{abstract}




\section{Introduction}

High-redshift ($z>4.5$) radio sources (HzRSs) provide an unparalleled dust-free window into the physics of galaxies in the early Universe.
Their radio activity originates from synchrotron radiation produced either by the release of energy by accreting supermassive black holes (SMBHs) in active galactic nuclei (AGN), or by short-lived massive stars that emit cosmic rays when turning supernova, providing a tracer of the star-formation rate. 
This means HzRSs are ideal laboratories for studying the physical mechanisms governing the growth and eventual quenching of high-redshift galaxies, such as gas accretion, black hole growth and feedback processes impacting the interstellar medium. 
Powerful HzRSs are also often seen at the centres of massive protoclusters, with a number of star-forming galaxies with strong Lyman-$\alpha$ and $\ha$ emission seen in their environments \citep[e.g.][]{Venemans07, Miley08, Muldrew15, Overzier16}.
Most HzRSs discovered to date are classified as ``radio-loud" AGN with radio flux densities only achieved by luminous jets and lobes produced by the accretion of material onto a central SMBH \citep[][]{Bornancini07, Bryant09, Jarvis09, Saxena18, Capetti25}.
However, below a luminosity of roughly $L_{\rm 1.3 \, GHz}\sim10^{24} \rm \, W \, Hz^{-1}$, the radio emission can arise either from AGN jets with lower powers than those in previous studies, or by intense star-formation \citep[SF, e.g.][]{Condon92, Sadler02, Mauch07, Novak_2017, Thykkathu2026}. This regime is largely unexplored by searches for HzRSs, due to the depth of the radio continuum data that is required to detect such sources at high redshift.
Both of these classes of HzRSs are interesting and can provide vital insights into rare sources at these redshifts; we are either detecting the most extreme starbursts at high redshifts (SFR $\sim10^2$--$10^3$ M$_\odot \, \rm yr^{-1}$) if all the radio emission is due to star formation \citep[e.g.][]{Algera20, Zavala23}, or a population of lower-powered radio jets which have not previously been studied at these redshifts, which could be responsible for shutting down star-formation at such early epochs \citep[e.g.][]{Dubois13}.
In the former case, the radio detections of these extreme starbursts will allow us to measure their SFRs unbiased by dust obscuration.
Moreover, radio observations provide an efficient method of identifying these rare objects due to the combination of areal coverage and depth, compared to ALMA which is limited to narrow fields \citep[e.g.][]{Franco18}, and far-infrared observations \citep[e.g. SCUBA, Herschel,][]{Smail02, Lapi11} which are heavily confusion limited.

Previous searches for HzRSs have been conducted by selecting relatively bright ($>1$\,mJy) ultra-steep spectrum (USS, $\alpha\geq1$, $S_{\nu} \propto \nu^{-\alpha}$) radio sources in shallow surveys covering several thousand square degrees  \citep[e.g.][]{Rawlings1996,DeBreuck2000,Jarvis2001,Saxena18_search, Capetti25, Ighina25}, limiting the source discovery space to luminous radio-loud AGN and quasars, to the extent that very few would be expected \citep{Jarvis2000,Jarvis2001c, Brookes2008,Ker2012}.
\citet{Jarvis09} dropped the USS requirement, instead cross matching radio sources which are very faint or not detected in \textit{Spitzer}/IRAC or $K$-band imaging, ensuring sources lie at $z>2$, leading to the discovery of a radio galaxy at $z=4.88$.
More recently, \citet{Endsley22} discovered a radio-loud AGN at $z=6.8$ by crossmatching deep VLA imaging with a sample of Lyman-break galaxy (LBG) candidates, which was later confirmed with ALMA \citep{Endsley23}, demonstrating the strength of cross matching optical+NIR-selected high-redshift galaxies with deep radio observations.
\citet{Lambrides24} have presented a candidate radio-loud AGN at $z=7.7$, but this source still requires spectroscopic confirmation.

A population of low-luminosity HzRSs may be relevant in the broader context of recently identified “Little Red Dots” (LRDs; \citealt{Matthee24}) discovered with \jwst.
LRDs are characterised by compact morphologies, distinctive `V-shaped' spectral energy distributions, and show evidence for AGN activity from broad rest-frame optical emission lines \citep[e.g.][]{Greene24}, with the AGN potentially enshrouded by a dense cocoon of gas \citep[e.g.][]{Naidu25bhstar, deGraaff25, Sun26}.
To date, however, no radio emission associated with accreting SMBHs has been detected from LRDs, including in stacking analyses \citep[][]{Mazzolari24, Perger25, Orozco25, Akins25}.
It therefore remains unclear whether their apparent radio-quiet nature reflects intrinsically weak radio emission compounded by current sensitivity limits, or instead points to alternative power sources, such as extremely compact and dense star-forming regions \citep{Baggen24, Guia24}.
Recent radio observations of a low-redshift ($z=0.17$) analogue to an LRD with the VLA suggest that at least some of these systems may host radio emission consistent with radio-luminous supernovae \citep{Rodriguez25}.
Faint HzRSs could provide complementary insight into high-redshift systems powered by either low-luminosity AGN or intense star formation, where the radio emission is detectable \citep{Fu25}.

By utilising deep radio imaging from the MeerKAT International GHz Tiered Extragalactic Exploration survey Data Release 1 \citep[MIGHTEE DR1;][]{Jarvis16, MIGHTEE_DR1} at 1.3 GHz, together with excellent multi-wavelength data, we are able to extend the study of high-redshift radio sources to lower-radio powers and larger areas for the first time.

This paper is structured as follows. 
In Section~\ref{sec:data} we present the multi-wavelength data and methods used to select the HzRSs.
We present the three spectroscopically confirmed HzRSs in Section~\ref{sec:results}, and compute their star formation rates from rest-frame UV, optical and radio tracers.
In Section~\ref{sec: discussion} we explore whether these HzRSs are star-formation or AGN-dominated, discuss their spectral indices, and compare to expected numbers from the radio luminosity function.
Finally, we present our conclusions in Section~\ref{sec: conclusions}.

All magnitudes reported in this paper are in the AB system \citep[][]{Oke83}.
We assume a standard cold dark matter cosmology, $H_0 = 70 \, \rm km \, s^{-1} \, Mpc^{-3}$,  $\Omega_{\rm{m}}=0.3$, $\Omega_{\Lambda}=0.7$.
We use a \citet{Kroupa01} initial mass function (IMF) throughout.
For radio spectral indices $\alpha$, we use the convention $S_{\nu}\propto\nu^{-\alpha}$ for a source with flux density $S_{\nu}$ at frequency $\nu$.
Errors are reported at $1\,\sigma$.

\section{Data and HzRS selection}
\label{sec:data}

In this Section~we present the multi-wavelength data and methods used to select the HzRSs.

\begin{figure}
    \centering
    \includegraphics[width=\linewidth]{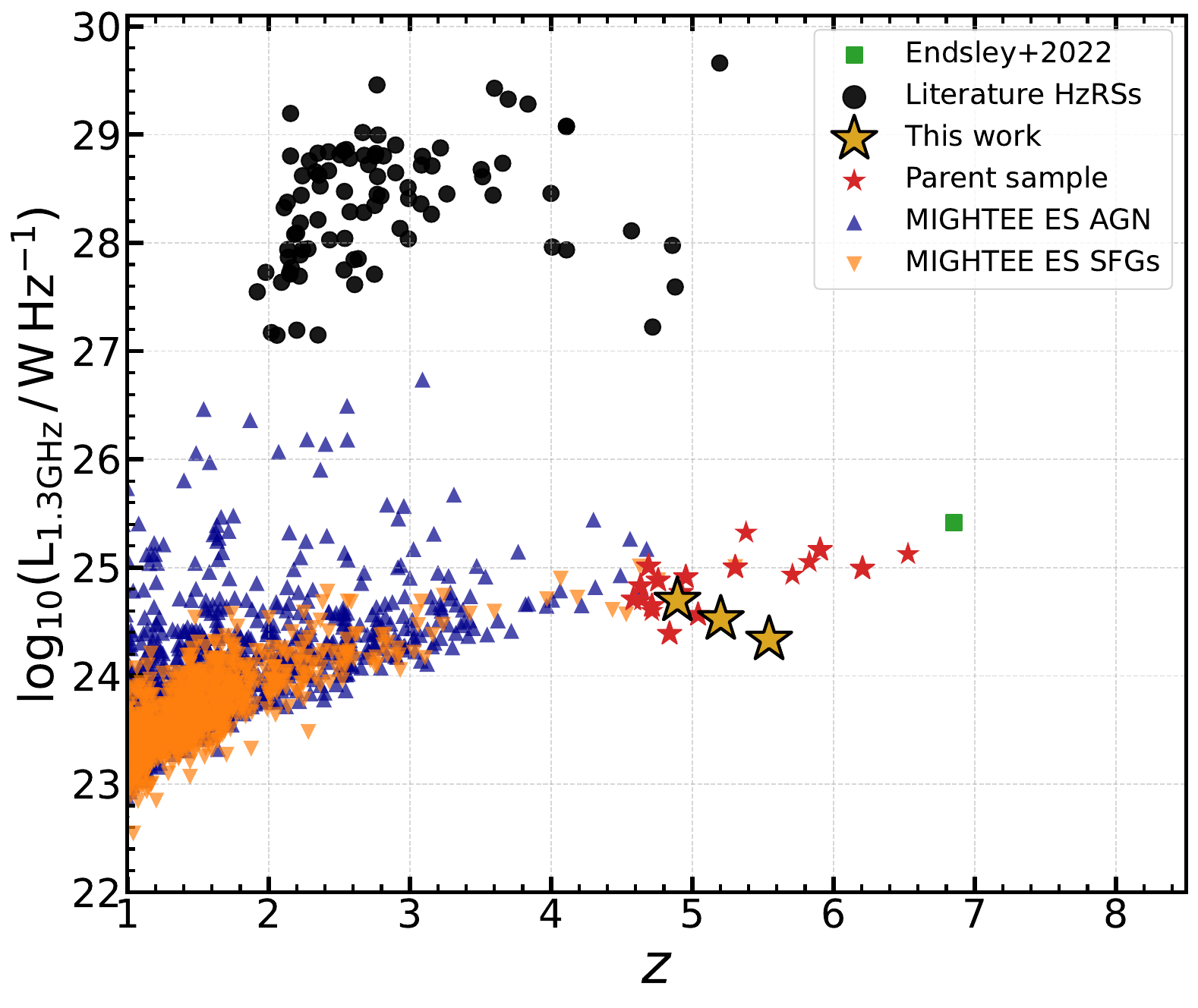}
    \caption{The radio luminosities of the spectroscopically confirmed HzRSs presented in this work, shown by the gold stars. We show the parent sample which overlaps with COSMOS-3D, determined from crossmatching MIGHTEE radio sources to sources with a photometric redshift $z_{\rm phot}>4.5$, by the smaller red stars.
    The coloured triangles represent AGN (blue) and SFGs (orange) from the MIGHTEE Early Science data \citep[MIGHTEE ES,][]{Whittam22}.
    The black points represent samples from the literature, compiled from \citet{Bornancini07, Bryant09, Jarvis09, Saxena18_search}, scaled to 1.3 GHz assuming a spectral index of $\alpha = 0.7$.
    The green square shows the highest-redshift confirmed radio loud AGN discovered by \citet{Endsley22}.
    }
    \label{fig:L_R vs redshift}
\end{figure}

\subsection{MIGHTEE Radio data}
\label{sec:mightee data}

The MIGHTEE survey is a MeerKAT \citep{Jonas2009} radio survey providing simultaneous continuum \citep{Heywood22, MIGHTEE_DR1}, spectral line \citep{Heywood2024} and polarisation \citep{Taylor2024} data.  This work is based on MIGHTEE Data Release 1 continuum data in the COSMOS field at 1.3 GHz, outlined in \citet{MIGHTEE_DR1}. 
The COSMOS data consist of 22 pointings, with a total on-target time of 139.6 h, covering $\rm 4.2 \, deg^2$. Two different versions of the imaging and associated source catalogues are produced, one optimised for sensitivity and one optimised for resolution \citep[see][for details]{MIGHTEE_DR1}. In this work we use the higher-resolution image, which has a circular synthesized beam full-width half maximum of 5.2 arcsec and the measured noise in the centre of the image is $2.4 \, \rm \mu Jy \, beam^{-1}$.

\begin{figure*}
    \centering
    \includegraphics[width=0.3\linewidth]{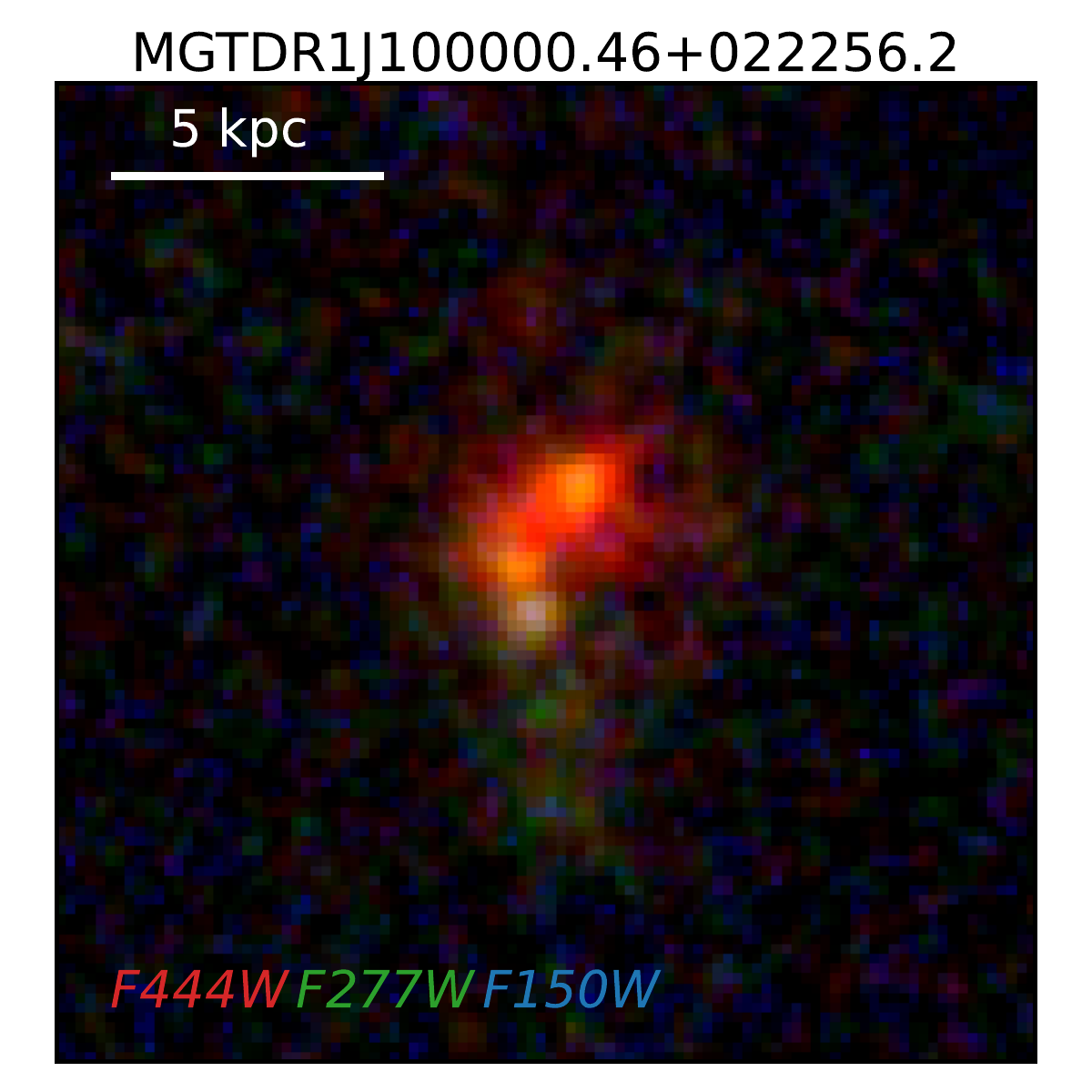}
    \includegraphics[width=0.3\linewidth]{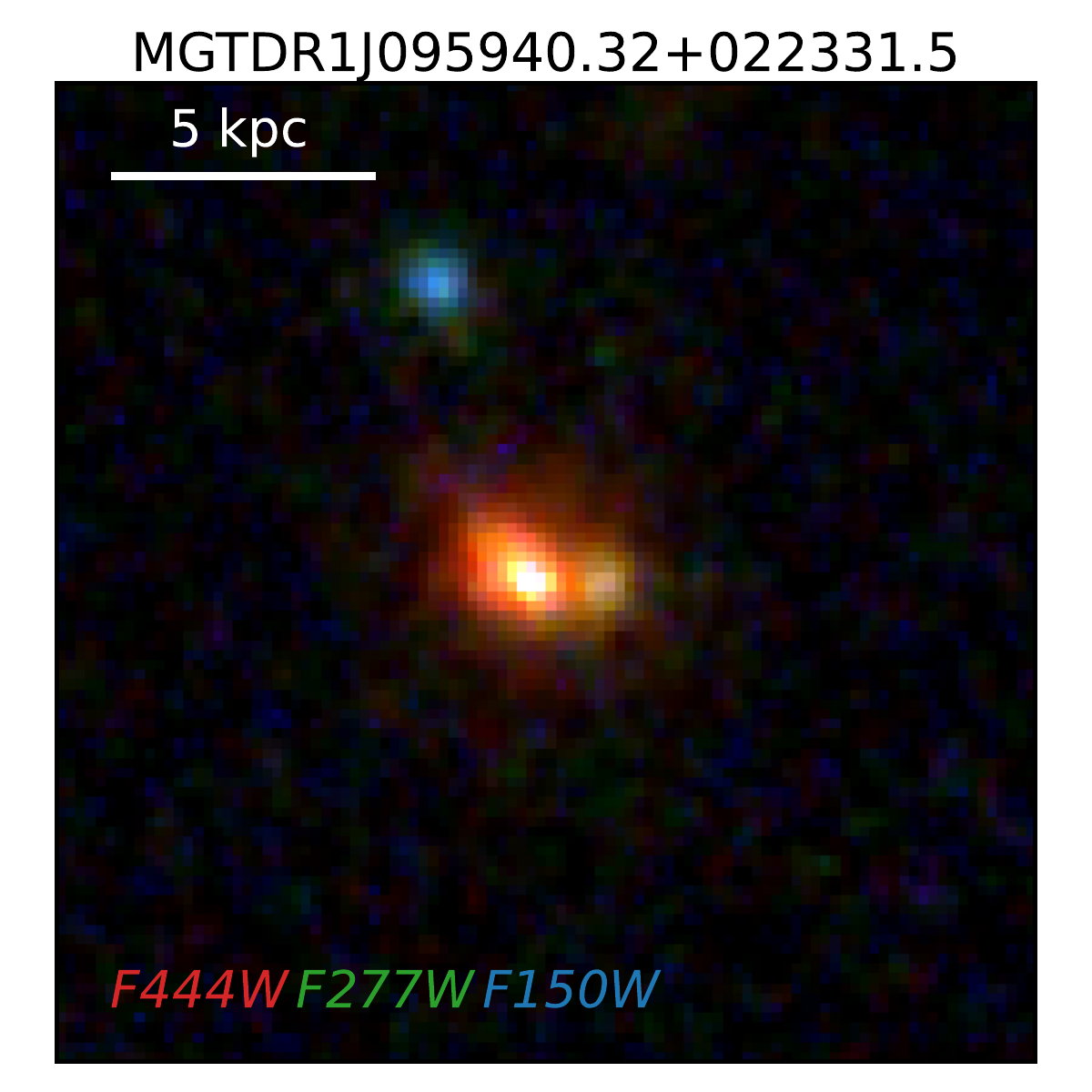}
    \includegraphics[width=0.3\linewidth]{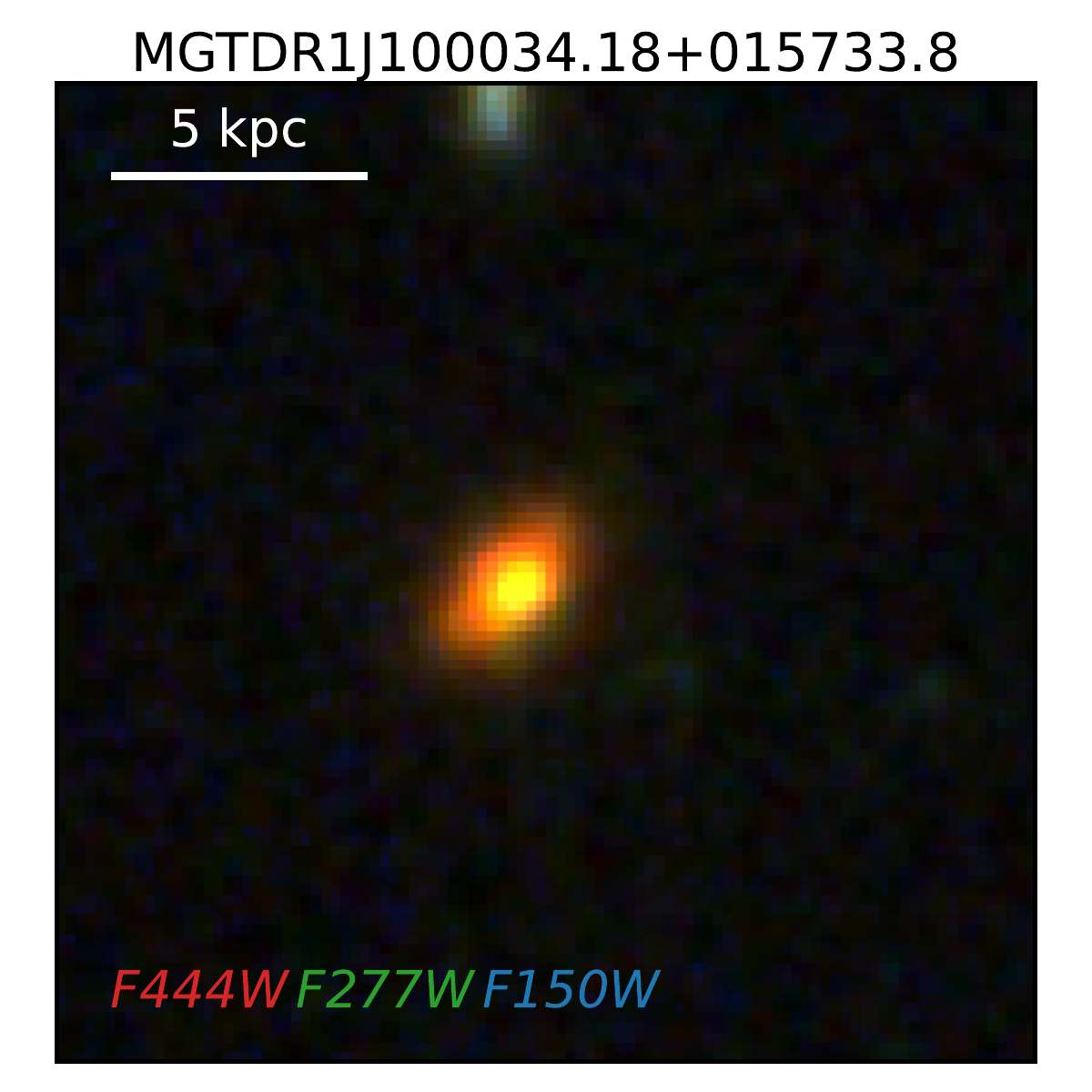}
    \caption{RGB images of the HzRSs presented in this work.
    The images are $3\times3$ arcsec, and are generated using \textit{F444W} (red), \textit{F277W} (green), and \textit{F150W} (blue). 
    A 5 kpc scale bar is shown for reference.
    We note that \agnshort and \sfgshort are composed of multiple components, whereas \starburstshort shows a disk-like morphology with a possible bulge.
    }
    \label{fig:rgb}
    
\end{figure*}

An earlier version of the MIGHTEE radio data, consisting of a single pointing in the COSMOS field \citep[the MIGHTEE Early Science data, see][for details]{Heywood22}, was crossmatched with deep ground-based optical and NIR imaging (see next section) in \citet{Whittam24}. We use this Early Science (ES) cross-matched catalogue, together with the deeper DR1 radio catalogue, as the starting point for this work as outlined in the following sections.
Due to the wide bandwidth of the MeerKAT $L$-band receiver the effective frequency of the MIGHTEE DR1 data varies slightly across the image. We use the effective frequency map released with the DR1 data to scale the radio flux density of each source to 1.3 GHz.
Fig.\ \ref{fig:L_R vs redshift} shows the redshift-- radio luminosity plane for the SFGs and AGN from the MIGHTEE Early Science (ES) data \citep{Whittam22} which extends to $z\approx4.5$.
~

\subsection{Ground-based data}
\label{sec:ground-based data}

We make use of rest-UV selected catalogues produced by \citet{Adams23} in the COSMOS field.
The full selection is summarised in their Section~2.
Deep optical data from the Canada-France-Hawaii-Telescope Legacy Survey \citep[CFHTLS;][]{CFHTLS} and the Hyper-Suprime Cam Subaru Strategic Program \citep[HSC-SSP;][]{HSCSSP_DR2} were combined with NIR data from the UltraVISTA survey \citep{UltraVISTA}, covering around $\rm 1.5\, deg^2$.
Sources in this catalogue with $K_s<25$ were crossmatched to the MIGHTEE Early Science catalogue using both the likelihood ratio method \citep{McAlpine2012} and visual inspection \citep{Whittam24}.
Photometric redshifts were determined using both a template-fitting method using \textsc{LePhare} \citep[][]{arnouts99, ilbert06} and machine learning with \textsc{GPz} \citep[][]{Almosallam16a, Almossalam16b, Hatfield20, Hatfield22}.
Less than $5$ per cent of sources have a photometric redshift in significant disagreement with available spectroscopic redshifts.
However, at $z>4$, spectroscopic redshifts become sparse, and a careful treatment of interloper sources is required.
\citet{Adams23} go on to select a sample of LBGs at $z\simeq3.5-5.5$ by using \textsc{LePhare} to split the catalogue into galaxies, AGN and Milky Way stars, mitigating contamination by, for example, ultra-cool dwarfs \citep[with M, L, and T-type templates from][]{burgasser14}.
We make use of their $z=5$ sample, which contains sources with $4.5<z_{\rm phot}<5.5$. 
We crossmatch these LBGs to the host galaxy positions from the \citet{Whittam24} MIGHTEE ES cross-matched catalogue.

\subsection{\jwst COSMOS-Web}
\label{sec:cosmos-web}

\begin{figure*}
    \centering
    \includegraphics[width=0.49\linewidth]{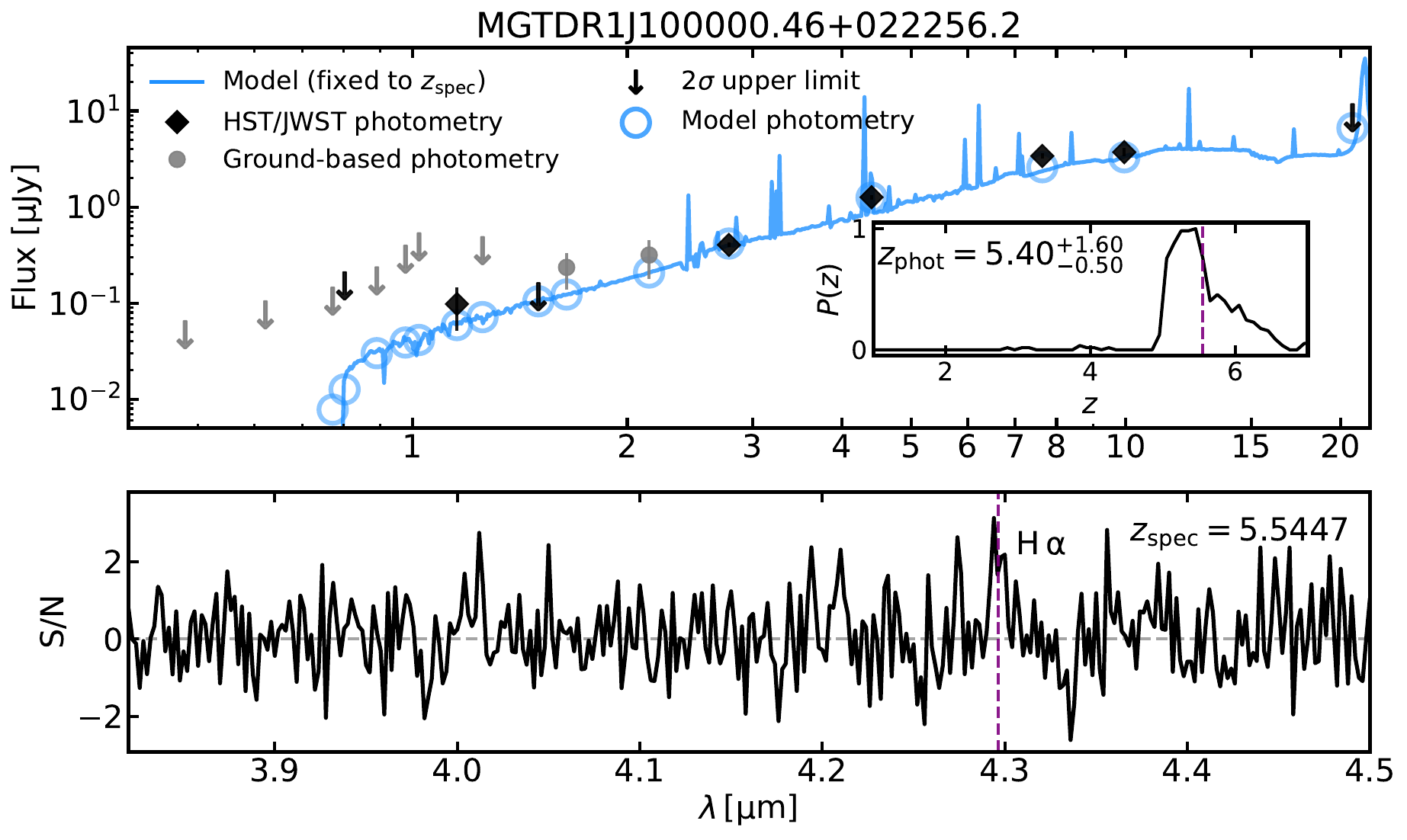}
    \includegraphics[width=0.49\linewidth]{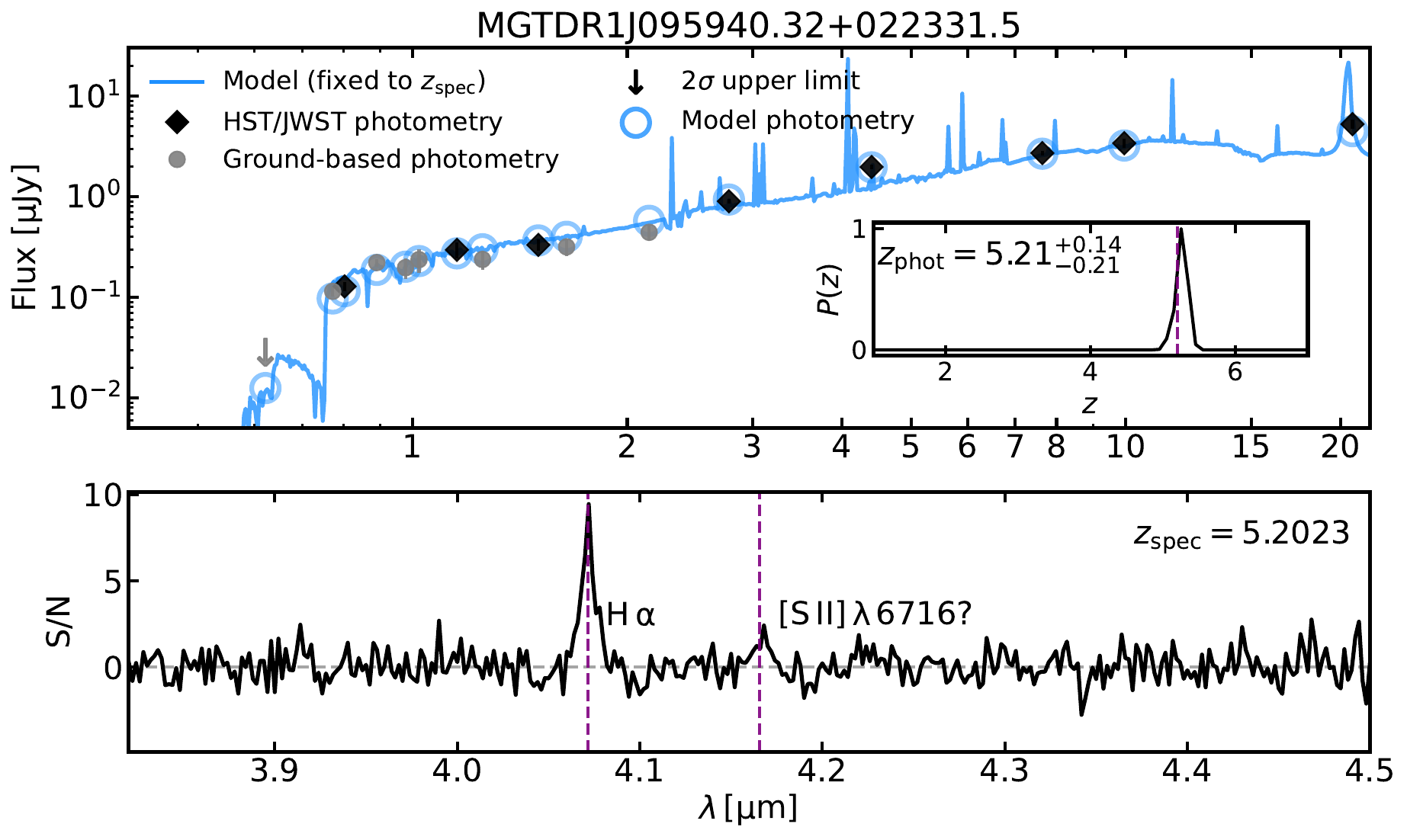}

    \includegraphics[width=0.5\linewidth]{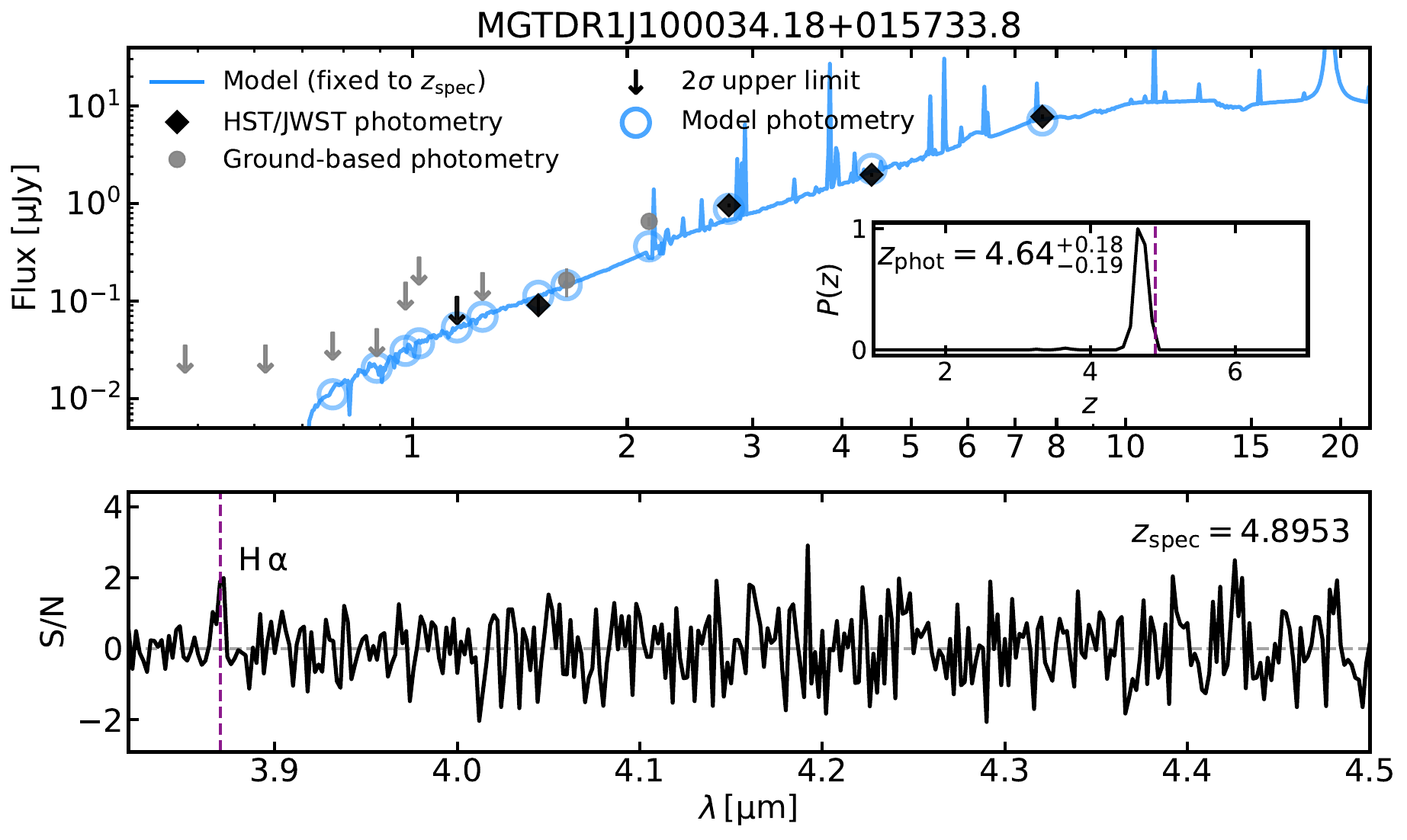}
    \caption{SED fitting (top panels) and \jwst grism spectroscopy (bottom panels) of the three HzRSs presented in this work.
    For each source, the top panel shows the best-fit SED from \textsc{BAGPIPES} when we fix the redshift to $z_{\rm spec}$.
    The black diamonds indicate the photometry from \textit{Hubble} F814W, \jwst NIRCam, and \jwst MIRI.
    The grey points indicate the ground-based photometry from Subaru HSC and VISTA.
    Non-detections are replaced with arrows indicating $2\,\sigma$ upper limits.
    Open blue circles indicate the model photometry.
    The inset panel shows the redshift probability distribution, $P(z)$, when we place a flat prior on the redshift instead of fixing it to $z_{\rm spec}$.
    The purple dashed lines in the bottom panels indicate the position of detected emission lines matching the photometric redshift.
    We also mark the corresponding spectroscopic redshift on the $P(z)$ panel. We find excellent agreement between the photometric redshifts derived from using flat priors and the spectroscopic redshifts measured from the grism spectra for all three sources.
    }
    \label{fig:SEDs}
\end{figure*}

Since the MIGHTEE catalogues are matched to ground-based imaging, it is entirely possible that highly reddened and intrinsically faint sources will be missed, leading to no ground-based counterpart.
We therefore also crossmatch the MIGHTEE catalogue to COSMOS2025 \citep{COSMOS2025}.
Briefly, this catalogue was constructed by running \textsc{SEP} \citep{SEP}, a Python implementation of \textsc{Source Extractor} \citep{sextractor}, on a $\chi^2_{+}$ stack of the \jwst NIRCam \textit{F115W}, \textit{F150W}, \textit{F277W}, and \textit{F444W} filters from the $\rm 0.54 \, deg^2$ COSMOS-Web survey \citep[][]{CWEB}.
\textsc{SourceXtractor++} \citep{Bertin20, Kummel20, Kummel22} is then used to perform multi-band model-fitting photometry across all available ground- and space-based imaging. 
\textit{Hubble} \textit{F814W} imaging \citep{Koekemoer07} is used to reject low-redshift interlopers, while \jwst MIRI \textit{F770W} data from COSMOS-Web provide improved stellar mass and dust constraints \citep[e.g.][]{Papovich23}. 
MIRI imaging covers 0.2 deg$^2$, such that only a subset of the 18 HzRS candidates (Section~\ref{sec:sed fitting}) have MIRI coverage.
SED fitting is then conducted with \textsc{LePhare} \citep{arnouts99, ilbert06}, with the outlier fraction compared to spectroscopic samples not exceeding 10 per cent down to $m_{F444W}= 28$.
We retain sources with $z_{\rm phot}\geq 4.5$, with this threshold chosen to extend beyond the MIGHTEE ES AGN/SFG separation carried out by \citet{Whittam22}.

\begin{table*}
    \centering
    \caption{Observed spectroscopic, photometric, emission-line, and radio properties of the HzRSs presented in this work, ordered by decreasing spectroscopic redshift.
    Source IDs are from the MIGHTEE DR1 catalogue.
    $z_{\rm spec}$ is determined from the $\ha$ line position in the NIRCam/WFSS data, while $z_{\rm phot}$ is obtained from broadband SED fitting.
    $f_{\ha}$ is the hobserved $\ha$ line flux density.
    $S_{1.3\,\rm GHz}$ and $S_{3\,\rm GHz}$ represent the flux densities at 1.3~GHz from MIGHTEE and 3~GHz from VLA-COSMOS \citep{Smolcic2017} respectively. MGT J09594+02233 and MGT J10000+02225 are undetected in the \citeauthor{Smolcic2017} catalogue, the 3~GHz flux densities listed for these sources are extracted from the radio image.
    $L_{\rm{1.3\,GHz}}$ is the rest-frame radio luminosity computed using the measured spectral index $\alpha$, while $L_{\rm{1.3\,GHz}}^{\alpha=0.7}$ assumes a fixed spectral index of $\alpha=0.7$, both using $z_{\rm spec}$.
    Finally, $\alpha^{\rm 3\, GHz}_{1.3\rm \, GHz} $ is the spectral index, measured from $S_{1.3\,\rm GHz}$ and $S_{3\,\rm GHz}$.
    }
    \begin{tabular}{l c c c c c c c c}
        \hline
        MIGHTEE DR1 ID  
        & $z_{\rm spec}$ 
        & $z_{\rm phot}$  
        & $f_{\ha}$ 
        & $S_{\rm 1.3 \, GHz}$
        & $S_{\rm 3 \, GHz}$
        & $L_{\rm{1.3\,GHz}}^{\alpha=0.7}$ 
        & $L_{\rm{1.3\,GHz}}$ 
        & $\alpha^{\rm 3\, GHz}_{1.3\rm \, GHz} $\\
        & 
        & 
        & [$10^{-17}$ erg/s/cm$^{2}$] 
        & [$\muup \rm Jy$]
        & [$\muup \rm Jy$]
        & [$10^{24}$ W/Hz] 
        & [$10^{24}$ W/Hz] \\
        \hline
        MGTDR1\,J100000.46+022256.2 
        & $5.5447$ 
        & $5.50^{+1.60}_{-0.50}$  
        & $1.03\pm0.30$ 
        & $11.5 \pm 3.5$
        & $2.6 \pm 2.2$
        & $2.16\pm0.67$ 
        & $13.6\pm18.4$ 
        &  $1.68 \pm 0.70$ \\
        MGTDR1\,J095940.32+022331.5 
        & $5.2023$ 
        & $5.21^{+0.14}_{-0.21}$  
        & $5.16\pm0.20$ 
        & $20.1 \pm 4.6$
        & $8.7 \pm 2.1$
        & $3.34\pm0.77$ 
        & $5.36\pm4.11$ 
        & $0.96 \pm 0.40$ \\
        MGTDR1\,J100034.18+015733.8  
        & $4.8953$ 
        & $4.64^{+0.18}_{-0.19}$  
        & $2.99\pm1.13$ 
        & $34.0 \pm 4.2$
        & $14.2 \pm 2.3$
        & $4.95\pm0.61$ 
        & $8.29\pm3.53$ 
        & $0.99 \pm 0.23$ \\
        \hline
    \end{tabular}
    \label{tab:properties}
\end{table*}

 \begin{figure*}
    \centering

    \includegraphics[width=\linewidth]{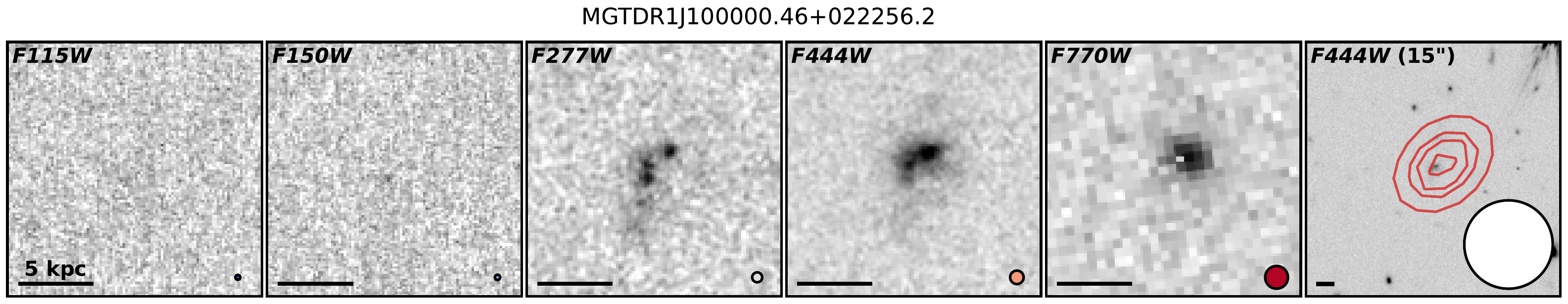}
    
    \includegraphics[width=\linewidth]{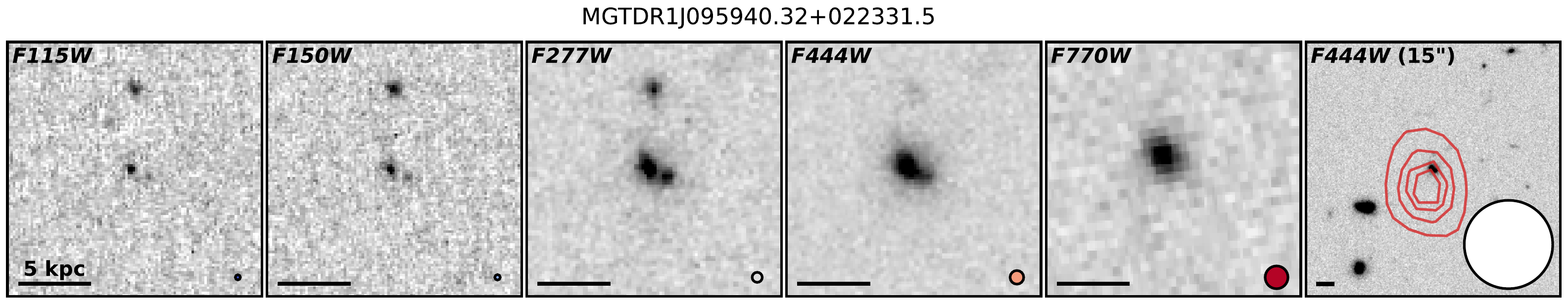}

    \includegraphics[width=\linewidth]{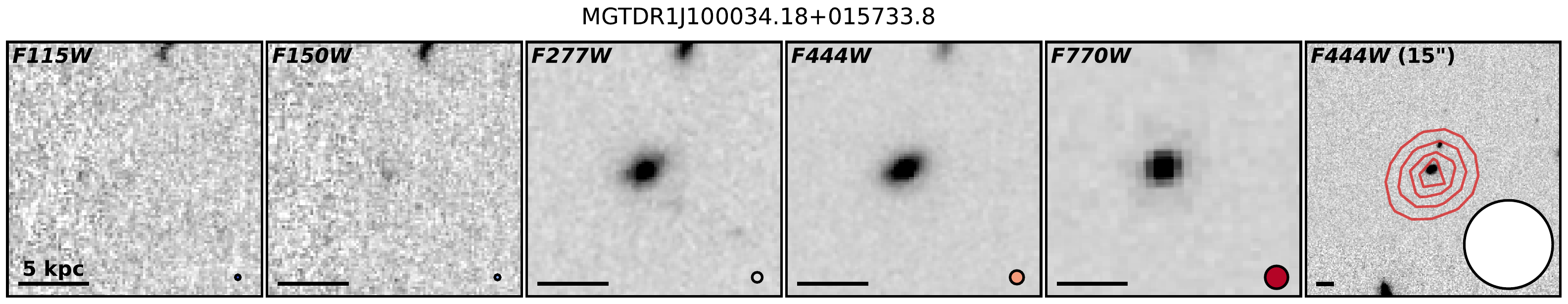}
    
    \caption{\jwst NIRCam postage stamp cutouts of the HzRSs presented in this work. 
    For each source, the first four stamps are $3\times3$ arcsec. 
    The final postage stamp cutout is $15\times15$ arcsec.
    In this final cutout the MIGHTEE continuum contours are overlaid.
    Contours are drawn at the 90th, 95th, 97.5th, and 99th percentiles, and at the maximum, of the radio cutout pixel values.
    The \jwst images saturate at $2\,\sigma$ below and $8\,\sigma$ above the noise level. 
    We show a 5 kpc scale bar in the bottom left, and the \jwst PSF FWHM and MIGHTEE beam size in the bottom right.
    Although the radio contours for \agnshort appear offset from the NIR position, the offset is $1$ arcsec, which is within the uncertainty of the radio centroid at the signal-to-noise ratio of the 1.3 GHz detection.
    }
    \label{fig:stamps}
\end{figure*}

\subsection{SED Fitting}
\label{sec:sed fitting}

We use SED fitting to identify robust HzRS candidates and remove low-redshift interlopers from the MIGHTEE-crossmatched sample.
Sources from the ground-based catalogue of \citet{Adams23} and from COSMOS2025 \citep{COSMOS2025} have already undergone SED fitting analysis using \textsc{LePhare} \citep{arnouts99, ilbert06}, with an outlier rate of less than 10 per cent across both studies, when compared to spectroscopic samples.
We crossmatch the ground-based candidates to the COSMOS2025 catalogue, obtaining consistent model-based photometry from \citep{COSMOS2025} for all available photometry from \jwst, \textit{Hubble}, VISTA, and Subaru Hyper Suprime-Cam (HSC).
We use \textsc{BAGPIPES} \citep{BAGPIPES} to verify the photometric redshifts from \textsc{LePhare}.
We use the non-parametric star formation history (SFH) described by \citet{Iyer19} with six lookback time bins and a Dirichlet prior on the lookback times.
We place flat priors on the age of the galaxy (100 Myr to the age of the Universe at the redshift), stellar mass ($10^6$M$_{\odot}$ to $10^{13}$M$_{\odot}$), dust attenuation ($A_V=0$--$4$ using the \citealt{Calzetti97} law), ionization parameter ($\log U = -4$ to $-2$), and redshift ($z=0$--$10$).
Candidate HzRSs are then taken as sources which have agreeing photometric redshifts at $z>4.5$ from both \textsc{LePhare}+\textsc{BAGPIPES} runs.
We find no significant differences when the shallower ground-based data is excluded.
The SED fitting analysis yields 18 HzRS candidates which overlap with the COSMOS-3D footprint (see Section \ref{sec:cosmos3d}).
We show their photometric redshifts as a function of their 1.3 GHz luminosities in Fig. \ref{fig:L_R vs redshift}, alongside HzRSs discovered previously, highlighting the very low radio luminosities we are probing with the new MIGHTEE data compared to previous studies.

\begin{table*}
    \centering
    \caption{Derived physical properties of the HzRSs.
    Dust attenuation ($A_V$) and stellar mass are obtained from SED fitting with \textsc{BAGPIPES}.
    Effective radius $R_{\rm e}$ and S\'ersic index $n$ are taken from S\'ersic profile fits in the COSMOS2025 catalogue \citep{COSMOS2025}.
    Star formation rates are inferred independently from SED fitting (averaged over 10 Myr and 100 Myr), dust-corrected $\ha$ emission, and radio luminosity at 1.3\,GHz assuming a \citet{Kroupa01} IMF.
    The SED fitting-derived properties are measured by fixing the redshift to the spectroscopic redshift, with a small prior ($z_{\rm spec}\pm0.05$).}
    \begin{tabular}{l c c c c c c c c}
        \hline
        MIGHTEE DR1 ID  
        & $A_V$ 
        & $\log_{10}(M/\rm M_\odot)$ 
        & $R_{\rm e}$ 
        & $n$
        & $\rm SFR^{SED}_{\rm 10 \, Myr}$ 
        & $\rm SFR^{\ha}$ 
        & $\rm SFR^{SED}_{\rm 100 \, Myr}$ 
        & $\rm SFR^{1.3\,GHz}$ \\
        & [mag] 
        & 
        & [kpc] 
        & 
        & [$\rm M_\odot$/yr] 
        & [$\rm M_\odot$/yr] 
        & [$\rm M_\odot$/yr]
        & [$\rm M_\odot$/yr]\\
        \hline
        MGTDR1\,J100000.46+022256.2  
        & $2.66^{+0.17}_{-0.18}$ 
        & $10.72^{+0.21}_{-0.23}$ 
        & $2.11\pm0.12$ 
        & $1.08\pm0.12$
        & $322_{-176}^{+161}$ 
        & $137\pm40$ 
        & $154_{-37}^{+49}$
        & $226\pm23$\\
        MGTDR1\,J095940.32+022331.5 
        & $1.64^{+0.07}_{-0.07}$ 
        & $10.16^{+0.11}_{-0.12}$ 
        & $1.36\pm0.03$ 
        & $1.66\pm0.08$
        & $300^{+50}_{-46}$ 
        & $273\pm32$ 
        & $75^{+16}_{-14}$ 
        & $366\pm36$ \\
        MGTDR1\,J100034.18+015733.8  
        & $3.72^{+0.14}_{-0.21}$ 
        & $10.78^{+0.10}_{-0.08}$ 
        & $0.90\pm0.01$ 
        & $1.41\pm0.04$
        & $1820^{+244}_{-321}$ 
        & $817\pm308$ 
        & $386^{+60}_{-72}$ 
        & $569\pm55$ \\
        \hline
    \end{tabular}
    \label{tab:SFR}
\end{table*}

\subsection{COSMOS-3D}
\label{sec:cosmos3d}

COSMOS-3D is a \jwst Cycle 3 Large Programme \citep[268h,][]{COSMOS3D} obtaining NIRCam wide-field slitless spectroscopy (WFSS) in \textit{F444W}, covering $0.33 \, \deg^2$.
The details of the grism extraction are described in \citet{Meyer25}; Feige et al. in prep.
WFSS images, 1D and 2D spectra are produced within \textsc{grizli} \citep{grizli}).
For this, the NIRCam/WFSS data is reduced using \textsc{grizli}\footnote{https://grizli.readthedocs.io/} version 1.12.11 and \textsc{jwst} pipeline version 1.18, and CRDS context pmap 1293.
Sources from COSMOS2025 with $m_{F444W}<27.5$ \citep{COSMOS2025} which overlap with the COSMOS-3D footprint are extracted, with this threshold allowing for the detection of faint objects with a single line and no continuum at the sensitivity limit of the grism data.
As outlined in the previous section, we find 18 HzRS candidates at $z>4.5$ which lie in the COSMOS-3D footprint and have $m_{F444W}<27.5$.
We use a script from the DAWN \jwst Archive (DJA)\footnote{https://dawn-cph.github.io/dja/blog/2025/05/16/simplified\_cutout\_wfss/} which measures redshifts by forward-modelling the spectrum with \textsc{grizli}.
The cross-dispersion profile was fit with a Gaussian. 
Continuum and contamination were modelled and subtracted using a flexible 31-knot spline. 
Emission lines were modelled with Gaussian line profiles assuming a velocity dispersion of 150 km s$^{-1}$. 
The fit returned the best-fitting 2-D model and redshift solution for $\ha$ within the relevant spectral window.
We measure the $\ha$ fluxes from the 1D grism spectrum. 
The flux density is converted from counts s$^{-1}$ to erg s$^{-1}$ \AA$^{-1}$ cm$^{-2}$ using the provided flat-field response. 
We identify the peak near the redshifted $\ha$ wavelength and integrate over the line pixels.
Uncertainties are derived from Monte-Carlo resampling of the spectrum.
We retain candidate HzRSs with $\rm S/N > 2.5$ (integrated) for a $\ha$ emission line in agreement with the photometric redshift $z_{\rm phot}$ within $2\,\sigma$. 
While this S/N is low in isolation, these detections are evaluated jointly with broadband SED fitting. 
This yields three robust HzRSs with the $\ha$ emission line at a redshift in agreement with the photometric redshift.
The remaining 15 sources are undetected in the grism spectroscopy.
The non-detections are likely related to the star formation timescales probed by $\ha$ and the 1.3 GHz luminosity respectively. 
This is discussed further in Section \ref{sec: agn or sfg}.

We also check the MIRI \textit{F1000W} and \textit{F2100W} imaging obtained as part of COSMOS-3D. 
Out of the three HzRSs presented in Section \ref{sec:results},
one source is detected in both filters, one in just \textit{F1000W}, and the remaining source is not covered by the imaging.
We measure fluxes from these images with a flexible Kron aperture using \textsc{SEP} \citep{SEP, sextractor}.
To measure the physical properties of the three confirmed HzRSs, we repeat the SED fitting outlined in Section \ref{sec:sed fitting}, but now with a narrow prior of $z_{\rm spec}\pm0.05$ on the redshift, as determined from the COSMOS-3D WFSS.
We present \jwst RGB images of the HzRSs in Fig. \ref{fig:rgb}, and their COSMOS-3D spectra and SEDs in Fig. \ref{fig:SEDs}.

\section{Results}
\label{sec:results}

We identify three low-luminosity HzRSs at $z>4.5$.
Spectroscopic confirmation has been provided by \jwst WFSS from the COSMOS-3D survey \citep{COSMOS3D}.
Due to the faint nature of these sources, none have a rest-frame optical continuum detection in the grism data.
We show the radio luminosities as a function of redshift in Fig.\ \ref{fig:L_R vs redshift} for the parent sample of 18 HzRS candidates (photometric redshifts), and for the three spectroscopically confirmed HzRSs.
We show RGB images of the HzRSs in Fig. \ref{fig:rgb}.
These sources have radio luminosities $L_{\rm 1.3 \, GHz}\lesssim5\times10^{24} \rm \, W \, Hz^{-1}$ (see Fig.\ \ref{fig:L_R vs redshift}), lying at least two orders of magnitude below previously identified HzRSs.
This sample represents the first time we are able to probe a regime where the radio emission is potentially dominated by SF, rather than AGN activity, at these redshifts for a sample selected in the radio band. 
In this Section we present each of the HzRSs, discuss potential low-redshift contamination, describe their morphology and measure their star formation rates.
In Table \ref{tab:properties} we present the redshifts and observable properties of the HzRSs such as their $\ha$ fluxes, radio flux densities and luminosities, and spectral indices.
In Table \ref{tab:SFR} we present their derived properties, such as dust attenuation, stellar mass, S\'ersic fit parameters, and star formation rates.
Properties derived from SED fitting are measured by fixing to the spectroscopic redshift with a small prior ($\Delta z=\pm0.05$).
In Fig.\ \ref{fig:SEDs} we show the SED fitting and grism spectroscopy of the HzRSs.
In Fig.\ \ref{fig:stamps} we show \jwst NIRCam postage stamp cutouts of the HzRSs, overlaid with MIGHTEE radio contours.
For clarity, we use abridged source IDs (\sfgshort, \agnshort, and \starburstshort) throughout the remainder of the text.
The full MIGHTEE DR1 IDs are listed in Table \ref{tab:properties}.
\subsection{\sfglong}

\sfgshort is selected from the COSMOS2025 catalogue \citep{COSMOS2025}, and is not detected in the ground-based HSC and VISTA data.
SED fitting suggests this source is a massive dusty galaxy with $\log_{10}(M/{\rm M}_{\odot})=10.72^{+0.21}_{-0.23}$ and $A_V=2.66^{+0.17}_{-0.18}$.
The redshift probability distribution has a broad peak, with $z_{\rm phot} = 5.40^{+1.60}_{-0.50}$, driven by non-detections at $\lambda\lesssim2 \, \micron$.
The spectroscopic redshift of $z_{\rm spec}=5.5447$, assuming the line is $\ha$, lies within the peak of the redshift probability distribution.
\sfgshort is detected in \textit{F1000W}, but lies in a low signal-to-noise region of the \textit{F2100W} imaging, so is undetected at $21 \, \micron$.
This source is also detected with SCUBA-2 at $850 \, \micron$ \citep[ID S2COS850 78,][]{SCUBA2}, with $S_{850\,\micron} = 4.8\pm1.0 \, \rm{mJy}$.
We use this flux to calculate the SFR from the warm dust emission in Section \ref{sec: SFR}.

\subsection{\agnlong}

\agnshort is selected from the ground-based LBG sample \citep{Adams23}.
A deep ground-based optical non-detection with HSC $g$ (see Fig. \ref{fig:SEDs}) reveals a strong Lyman-break in the original selection.
SED fitting finds this source has a stellar mass $\log_{10}(M/{\rm M}_{\odot})=10.16^{+0.11}_{-0.12}$ with moderate dust attenuation, $A_V=1.64^{+0.07}_{-0.07}$. 
This source is detected across all available ground- and space-based filters redwards of $0.7 \, \micron$, including \textit{F1000W} and \textit{F2100W}, providing strong photometric constraints.
The photometric redshift of $z_{\rm phot}=5.21^{+0.14}_{-0.21}$ is in excellent agreement with the spectroscopic redshift of $z_{\rm spec}=5.2023$.
This source has the most robust detection of $\ha$ at $25.8\,\sigma$ integrated across the line, and there is also a possible detection of \textsc{[S\,ii]} $\lambda \, 6716$.
On first inspection, the $\ha$ appears to be broad.
We fit a narrow single Gaussian, narrow+broad double Gaussian, and Lorentzian profile to the emission line.
At the depth of the COSMOS-3D WFSS, we find no evidence for a preferred narrow+broad profile.
Instead, a Lorentzian is moderately preferred over a single Gaussian profile ($\Delta \mathrm{BIC}=5.7$, where BIC is the Bayesian Information Criterion).

\subsection{\starburstlong}

\starburstshort is selected from the MIGHTEE ES sample, with a photometric redshift $z_{\rm phot}>4.5$ determined with \textsc{LePhare}.
SED fitting suggests this source has the largest dust attenuation in the sample with $A_V=3.72^{+0.14}_{-0.21}$ and a large stellar mass $\log_{10}(M/{\rm M_{\odot}})=10.78^{+0.10}_{-0.08}$.
The $\ha$ spectroscopic redshift of $z_{\rm spec}=4.8953$ lies within the peak of the redshift probability distribution.
This source is not covered by \textit{F1000W} and \textit{F2100W} imaging from COSMOS-3D.

\subsection{Low-redshift emission line contamination}

\sfgshort and \starburstshort are highly reddened sources, which prevents robust constraints on the position of the Lyman break and results in non-detections in most ground-based filters. 
Consequently, their photometric redshift probability distributions are broad and exhibit secondary peaks at $z<4.5$. 
Paschen-$\alpha$ ($\rm Pa \, \alpha$) emitters at Cosmic Noon \citep[e.g.][]{Seymour26} are a potential source of contamination. 
If the detected emission lines in \sfgshort and \starburstshort were $\rm Pa \, \alpha$ ($\lambda_{\rm rest}=1.875\,\micron$), the implied redshifts would be $z_{\rm Pa\,\alpha}=1.29$ and $1.04$, respectively. 
However, there is negligible photometric redshift probability density at $z\approx1$--$1.3$ for either source, disfavouring this interpretation. 
For \sfgshort, the integrated probability below $z=4.5$ is $P(z<4.5)=0.016$, while for \starburstshort\ it is $P(z<4.5)=0.022$. 
Other emission-lines (e.g.\ He\,\textsc{i}, O\,\textsc{i}, the Ca\,\textsc{ii} triplet, or the [S\,\textsc{iii}] doublet) are possible at $z\sim3$--4, but these redshifts contribute only a small fraction of the total $P(z)$. 
We therefore adopt the high-redshift solutions, identifying the detected lines as $\ha$.

\subsection{Morphology}
\label{sec: morphology}

All three HzRSs presented in this work are fit with a S\'ersic profile in the NIRCam filters by \citet{COSMOS2025}.
Their effective radii and S\'ersic indices are presented in Table \ref{tab:SFR}.
All sources have S\'ersic indices consistent with a disk ($n\lesssim1.5$). 
The effective radii are consistent with size-mass relations at $z\simeq5$ \citep[e.g.][]{Varadaraj24, Allen25}.
They therefore do not have ultra-compact morphologies in the rest-frame optical like LRDs, which are generally unresolved \citep[e.g.][]{Akins25}. 
Instead, they are consistent with star-forming galaxies at $z\gtrsim5$ \citep[e.g.][]{Kartaltepe23}.

In addition to the RGB images of the HzRSs in Fig.\ \ref{fig:rgb}, we also show \jwst NIRCam and MIRI \textit{F770W} postage stamp cutouts in Fig.\ \ref{fig:stamps}, along with an overlay of the radio continuum contours.
\sfgshort appears to have an irregular morphology in \textit{F277W}, comprised of three distinct clumps possibly surrounded by diffuse emission.
\agnshort also shows two clumps, with one brighter than the other, in \textit{F115W} and \textit{F150W}, and consistent with the FWHM of the PSF in these filters.
These could either be due to unobscured AGN emission, or due to compact star formation.
Compact rest-frame UV morphologies at high redshift can coincide with strong high-ionization lines such as \textsc{N\,iv]} $\rm \lambda\,1486$, suggesting either strong compact starbursts or AGN activity \citep{Harikane25}.
The multiple components may also be evidence for ongoing mergers, but confirming this scenario is not possible without follow-up observations which probe the kinematics of the clumps.
We discuss further whether these sources are powered by star formation or AGN in Section~\ref{sec: agn or sfg}, but we note that follow-up observations with \jwst/NIRSpec of rest-frame UV emission lines would provide valuable insight into the nature of these sources.
We also note that none of the HzRSs are resolved in the MIGHTEE imaging (see Fig. \ref{fig:stamps}, where the radio contours are consistent with the beam size), which is expected at these redshifts for both SFGs and AGN, and therefore cannot be used to distinguish between the two.

\citet{COSMOS2025} also fit bulge+disk profiles to the sources.
This fitting is less reliable for \sfgshort and \agnshort due to their irregular, multi-component morphologies.
However, we note that in Fig. \ref{fig:rgb}, it appears that \starburstshort hosts a bluer bulge and a redder disk.
\starburstshort has a bulge-to-total ratio of 0.15 in \textit{F277W} and 0.22 in \textit{F444W}.
The bulge has a magnitude $m_{F444W}=24.8$, compared to $m_{F444W}=23.4$ for the disk.
\starburstshort therefore hosts a faint bulge component, which does not dominate over the total flux, but may host an AGN that contributes some fraction to the total flux.
Similarly, we note that neither of \sfgshort or \agnshort are dominated by a point-source component in Fig.\ \ref{fig:rgb}.
We therefore cannot rule out the contribution of an AGN to the appearance of these HzRSs (even for the compact rest-UV morphologies seen for \agnshort), but conclude that they do not dominate the rest-frame UV and optical emission.

\subsection{Star formation tracers}
\label{sec: SFR}

We compute SFRs based on the radio and $\ha$ luminosities, and from SED fitting.
We assume that the radio flux is powered entirely by star formation.
The radio star formation rate is computed following \citet{Delhaize17}, who provide a redshift-dependent calibration
\begin{equation}
    \mathrm{SFR} \, [\mathrm{M_{\odot}/yr}] = f_{\mathrm{IMF}}10^{-24}10^{q_{\mathrm{TIR}}(z, \alpha)}L_{\mathrm{1.4 \, GHz}} \, [\mathrm{W / Hz}]
\end{equation}
where $f_{\rm IMF}=1.063$ for a Kroupa IMF \citep{Madau14}, $L_{\rm 1.4 \, GHz}$ is the radio luminosity at 1.4 GHz, and $q_{\rm TIR}$ is the infrared-to-1.4 GHz radio luminosity ratio, which was found to evolve with redshift as $q_{\rm TIR}(z) = (2.88\pm0.03)(1+z)^{-0.19\pm0.01}$. 
We assume a spectral index $\alpha=0.7$ and correspondingly scale the 1.3 GHz luminosities to 1.4 GHz.
We note that whilst we have spectral index measurements for these sources (see Section \ref{sec: spectral indices}), these have very large errors, leading to large errors on the radio luminosity (see Table \ref{tab:properties}).
We therefore fix the spectral index to $\alpha=0.7$ to provide a direct comparison to other studies which follow the same methodology.

We compute the $\ha$-inferred SFRs following \citet{Hao11, Murphy11, Kennicutt12}, which is given by
\begin{equation}
    \log_{10}(\mathrm{SFR/M_{\odot}\,yr^{-1}}) = \log_{10}(L_{\ha}/\mathrm{erg\,s^{-1}}) - 41.27
    \end{equation}
where $L_{\ha}$ is the $\ha$ luminosity.
In order to derive a dust-corrected $\ha$ luminosity, we assume that $A_{V, \rm stars} = A_{V, \rm gas}$, i.e. both the stars and emission lines lie within stellar birth clouds, which has been shown to be a reasonable assumption at high redshift \citep{Shivaei15,Smit16}.
We take $A_V$ as measured from the SED fitting.
Follow-up observations with \jwst/NIRSpec would provide measurements of $\rm H\,\beta$, allowing for a dust correction that does not rely on SED fitting.

\begin{figure}
    \centering
    \includegraphics[width=\linewidth]{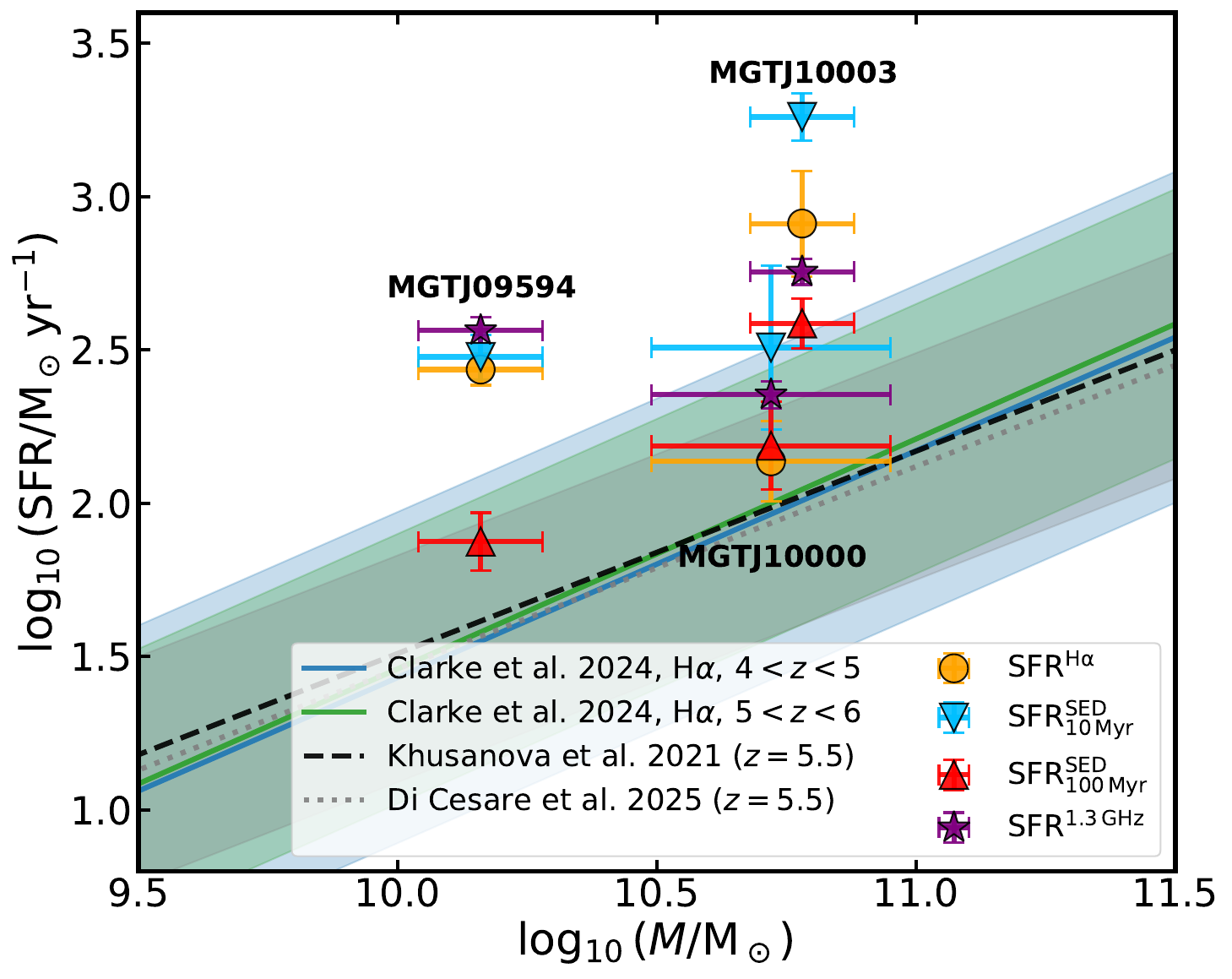}
    \caption{A comparison of the SFRs and stellar masses of the sources presented in this work against the star-forming main sequence at $z\simeq4-6$, compiled from \citet{Khusanova21, Clarke24, DiCesare25} and shown by the dashed, solid, and dotted lines respectively.
    The shaded regions indicate the intrinsic scatter.
    SFRs derived from the $\ha$ flux, SED fitting, and 1.3 GHz radio continuum (assuming Chabrier and Salpeter IMFs) are shown as red circles, blue squares, purple upward triangles, and yellow downward triangles, respectively.
    Each source is labelled by its abridged ID.
    }
    \label{fig:MS}
\end{figure}

Finally, the SED-inferred SFR comes from SED fitting with \textsc{BAGPIPES}, as described in Section~\ref{sec:sed fitting}, using a non-parametric star formation history.
We compute SFRs averaged over 10 Myr and 100 Myr ($\rm SFR^{SED}_{10\, Myr}$ and $\rm SFR^{SED}_{100\, Myr}$).
This is done to provide two SFR estimates which probe star-formation timescales comparable to $\ha$ ($\sim10$ Myr) and 1.3 GHz luminosity \citep[$\sim100$ Myr,][]{Kennicutt12}
We present the SFRs in Table \ref{tab:SFR}.
In Fig.\ \ref{fig:MS} we show the positions of the different SF tracers relative to the SF main sequence at $z\simeq4$--$6$.
We find that $\rm SFR^{SED}_{10\, Myr}$ is in excess of $\rm SFR^{SED}_{100\, Myr}$ for all sources, caused by a recent starburst in the SFH.
The SFRs measured from $\ha$ and the 1.3 GHz luminosity are either consistent with, or lie between, the $\rm SFR^{SED}$ values. 
Overall, the SFR values lie on, or up to 1 dex above, the SF main sequence, consistent with star-forming galaxies and starbursts.

We also convert the SCUBA-2 detection at $850 \, \micron$ for \sfgshort into a sub-mm derived SFR, following equation 2 from \citet{Carilli99}.
This detection probes emission at $130\,\micron$ in the rest frame at $z=5.5447$, tracing the warm dust.
Assuming a sub-mm spectral index $\alpha_{\rm sub-mm}=3.5$, we find $\mathrm{SFR^{850 \, \micron}}=330\pm69 \, M_{\odot} \, \mathrm{yr}^{-1}$, consistent with $\rm SFR^{SED}_{10\, Myr}$ and $\rm SFR^{1.3 \, GHz}$.
We note that observations with ALMA would also provide measurements of the warm dust.
Correspondingly, we crossmatch the HzRSs presented in this work with A$^3$COSMOS, which is a compilation of public Atacama Large Millimeter Array (ALMA) continuum data in the COSMOS field in bands 3-9 \citep[with large variance in area covered between bands,][]{Adscheid24}. 
However, we find no counterparts to our HzRSs due to a lack of coverage.

\subsection{Radio spectral indices}
\label{sec: spectral indices}

In Table \ref{tab:properties} we present the radio spectral indices of the HzRSs between 1.3 and 3~GHz.
\starburstshort is detected in the \citet{Smolcic2017} VLA 3~GHz data and has a radio spectral index $\alpha^{3~\mathrm{GHz}}_{1.3~\mathrm{GHz}} = 0.99 \pm 0.23$, indicating that it has a steep spectral shape. 
The other two sources are undetected in the \citeauthor{Smolcic2017} catalogue, but we measure their 3-GHz flux densities by extracting pixel values from the radio image. 
We find \agnshort has a radio spectral index $\alpha^{3~\mathrm{GHz}}_{1.3~\mathrm{GHz}} = 0.96 \pm 0.40$ and \sfgshort has $\alpha^{3~\mathrm{GHz}}_{1.3~\mathrm{GHz}} = 1.68 \pm 0.70$. 
Although the resulting spectral index estimates have large uncertainties, it is clear that both sources also have steep spectral shapes. 
The \citeauthor{Smolcic2017} observations lack sensitivity on short baselines meaning that some extended emission may be resolved out, causing an underestimation of the 3-GHz flux densities relative to the measured MIGHTEE flux densities \citep[see][]{Hale2023}. 
However, as the three sources in question are all compact, we do not expect this to be contributing to the measured steep spectral shapes. 
The radio spectral shapes of these sources are discussed further in Section~\ref{sec: spectral index discussion}.

\section{Discussion}
\label{sec: discussion}

In this Section we discuss whether the radio emission is powered by SF or AGN activity.
We then discuss the impact inverse-Compton scattering may have on our interpretation, and the expected number of radio sources based on the radio luminosity function.

\subsection{AGN or star formation?}
\label{sec: agn or sfg}

As shown in Fig.\ \ref{fig:L_R vs redshift}, the HzRSs presented in this work straddle a radio luminosity range on the boundary between SFGs and AGN, as determined by \citet{Whittam22}. 
A natural question to ask is whether we can determine whether these sources populate the high-SFR end of the high-redshift galaxy population, or if they are instead powered by low-luminosity AGN.
This requires use of both the rest-frame UV/optical morphology and SED fitting of these sources.
In Section \ref{sec: morphology}, we found that these sources are likely not dominated by AGN in the rest-frame UV and optical, based on bulge+disk fitting and their extended, clumpy nature.
Instead, we suggested that two of these sources may be undergoing mergers, evidenced by complex multi-component morphologies.

We can also use the results of SED fitting conducted by \citet{COSMOS2025}, who fit both AGN and galaxy templates to sources in the COSMOS2025 catalogue with \textsc{LePhare}.
\starburstshort has a marginally preferred galaxy template, with $\Delta\chi^2(\rm AGN -galaxy)=1.5$.
\sfgshort and \agnshort both have a preferred AGN template, with  $\Delta\chi^2(\rm galaxy-AGN)=3.5$ and 5.3 respectively.
Only \agnshort shows moderate preference for an AGN template at $2.3\,\sigma$, while the other two show no significant differences.
We note that distinguishing between these templates is limited by a lack of photometry.
However, we can again conclude that the SEDs of these sources are not dominated by an AGN.

In Fig. \ref{fig:SEDs}, a possible detection of \textsc{[S\,ii]} can be seen for \agnshort.
Assuming the \textsc{[S\,ii]} doublet is blended, we estimate $\log_{10}(\mathrm{\textsc{[S\,ii]}}/\ha)\sim-1.05$, but we note that the the detection of \textsc{[S\,ii]} is marginal, $f_{\rm \textsc{[S\,ii]}}=(4.5\pm6.6)\times10^{-18} \, \mathrm{erg \, s^{-1} \, cm^{-2}}$.
This value is consistent with both SFGs and AGN at high redshift \citep[e.g.][]{Scholtz25}, but without measurements of the $\rm \textsc{[O\,iii]}/H\,\beta$ ratio it is not possible to study this source further with \citet[][BPT]{BPT} diagnostics, used to identify ionisation mechanisms within galaxies and to distinguish SFGs from AGN.

Next we can compare the star-formation tracers discussed in Section~\ref{sec: SFR}.
For the purpose of this, we assume that the radio emission is driven entirely by star formation when measuring $\rm SFR^{1.3 \, GHz}$. 
Additionally, we note that BAGPIPES does not account for AGN emission.
We find broad agreement between all SFR tracers, shown in Fig.~\ref{fig:MS} against the star-forming main sequence at $z=4$--$6$.
For each source, $\rm SFR^{SED}_{10 \, Myr}$ is higher than $\rm SFR^{SED}_{100 \, Myr}$, suggesting that these galaxies are undergoing starbursts which are driving the $\ha$ emission.
We note that there is large scatter in the measured SFRs.
However, the errors shown in Fig.~\ref{fig:MS} and reported in Table \ref{tab:SFR} are only the statistical errors which do not account for systematics.
Both \sfgshort and \starburstshort have limited ground-based photometry, limiting the robustness of SED fitting-derived properties.
Their dusty nature results in poor constraints on the SFR from UV tracers, probing timescales of 100 Myr.
Additionally, different assumed SFHs can lead to scatter in the resulting SFR by $\sim 1$ dex \citep[e.g.][]{Tacchella22, Donnan26}.
Our dust correction of the $\ha$ flux is also based on SED fitting, rather than a more reliable Balmer decrement.
A related caveat to the lack of a measured Balmer decrement is provided by \citet{Ismail26}, who show that $\ha$-derived SFRs are susceptible to scatter from variations in the optical depth of dust along the line of sight.
Our results therefore point broadly to consistent SFRs across the three tracers, with these HzRSs lying on or up to 1 dex above the star-forming main sequence, identifying them as dusty star-forming galaxies.
Combining this, the morphology of the sources, and the fact that $\rm SFR^{SED}_{10 \, Myr}$ is higher than $\rm SFR^{SED}_{100 \, Myr}$, suggests that these galaxies are undergoing a starburst, allowing for the detection of the $\ha$ emission line.
The morphology of \agnshort and \sfgshort suggest they are merger-induced starbursts.
An AGN contribution cannot be ruled out, but is not likely to dominate over emission from star formation.

The interpretation of radio emission from starbursts may also explain the detection rate of $\ha$.
In Section~\ref{sec:cosmos3d} we found 18 photometric HzRS candidates which overlapped with the COSMOS-3D footprint, of which three are detected in $\ha$.
Since $\ha$ is sensitive to recent star formation on $\sim10$ Myr timescales, whereas GHz radio emission traces star formation averaged over $\sim100$ Myr, then if the radio emission is star-formation–driven, only $\sim10$ per cent of radio-selected sources are expected to exhibit $\ha$ emission at a given epoch (ignoring dust attenuation and selection effects).
This is consistent with our detection rate of $0.17^{+0.12}_{-0.06}$, with binomial errors determined from \citet{Cameron11}.

\subsection{Radio spectral shapes and inverse-Compton scattering}
\label{sec: spectral index discussion}

In Section \ref{sec: spectral indices} we measured the radio spectral indices of the HzRSs by combining VLA data at 3 GHz \citep{Smolcic2017} with MIGHTEE data at 1.3 GHz.
Although the uncertainties on the individual measurements are large as two of the sources are undetected in the 3-GHz catalogue, all three sources exhibit steep ($\alpha\gtrsim0.7$) radio spectral shapes.
The correlation between radio spectral steepness and redshift has been known for several decades \citep[e.g.][]{Tielens79} and has motivated the use of ultra–steep-spectrum (USS, $\alpha \gtrsim 1$) selection to identify HzRSs \citep{Rawlings1996,DeBreuck2000,Jarvis2001,Saxena18_search, Capetti25, Ighina25}. 
Some early interpretations attributed this trend to the dense environments of high-redshift radio galaxies, which may reduce fluid velocities and enhance radiative ageing, leading to steeper synchrotron spectra \citep[e.g.][]{Klamer06}. 
A more comprehensive overview of these scenarios is given by \citet{Miley08}.
However, the main physical mechanism expected to drive steep radio spectra at high redshift is energy loss of relativistic electrons via inverse-Compton (IC) scattering against cosmic microwave background (CMB) photons. 
While negligible in the local Universe, IC losses become increasingly important at high redshift because the CMB energy density scales as $(1+z)^4$, and are expected to dominate synchrotron losses at $z \gtrsim 3$ \citep{Murphy09}. 
Observational evidence for this effect was provided by \citet{Whittam25}, who stacked MIGHTEE observations of $\sim2\times10^5$ star-forming Lyman-break galaxies at $z=3$–5 and showed that, at fixed rest-frame UV magnitude, the flux density and luminosity at 1.4 GHz decline with redshift in agreement with predictions from \citet{Murphy09}.
The steep ($\alpha \gtrsim 0.7$) radio spectra observed for the HzRSs presented in this work can therefore be explained by cooling of relativistic electrons via IC scattering against the CMB, with any potential environmental effects playing a secondary role. 
If IC losses are affecting the radio emission from these sources, then this would cause the SFRs estimated from the radio emission in this work to be underestimates. Using the estimated $q_\textrm{TIR} \sim 2.4$ at $z=5$ found by \citet{Whittam25} instead of the redshift-dependent \citeauthor{Delhaize17} $q_\textrm{TIR}$ relation given in Section~\ref{sec: SFR} increases the radio SFRs by a factor of $\sim 2$. 
For \sfgshort and \starburstshort, this would still provide agreement between $\rm SFR^{1.3 \, GHz}$ and $\rm SFR^{SED}_{10 \, Myr}$.
While $\rm SFR^{SED}_{100 \, Myr}$ probes the SFR on a similar timescale to the 1.3 GHz luminosity, due to the scatter on the inferred SFRs from SED fitting, we can still conclude that there is no significant radio excess when IC losses are accounted for.
\agnshort would have $\rm SFR^{1.3 \, GHz}$ roughly a factor of two higher than its largest other SFR measurement, namely $\rm SFR^{SED}_{10 \, Myr}$.
While this could be interpreted as a radio excess, it is also consistent with a dusty starburst, which may not be well-constrained due to a lack of measurements of the warm dust emission, which reprocesses the UV light.
Follow-up observations with ALMA to measure the dust emission would break this degeneracy.
Additionally, deeper high-frequency radio data would reduce the uncertainties on the spectral indices and thus confirm this steepening, and this link to IC losses. We also note that determining the relationship between SFR and radio continuum luminosity based on full-SED modelling may also provide slightly lower SFRs than we find using the \cite{Delhaize17} relation. As demonstrated by \cite{Thykkathu2026}, adopting the more recent relation between SFR and radio luminosity from \cite{Cook2024} results in better agreement between the radio-derived and the combined UV+FIR measurements of the cosmic star-formation rate density.

\subsection{Expected Sources from the Radio Luminosity Function}
\label{sec: RLF discussion}

We are able to determine whether the radio sources we find at these high redshifts are consistent with the current best-fit radio luminosity function models from deep field data. 
The two most appropriate luminosity function evolution models at these faint flux densities come from the VLA-3GHz survey of \cite{Smolcic2017} and the MIGHTEE survey itself. 
\cite{Novak_2017} modelled the evolution in the star-forming galaxy population based on the VLA-3GHz data, constraining the bright-end ($L_{1.4} > 10^{24.5}$\,W\,Hz$^{-1}$) of the luminosity function at $z\sim 5$, although it is worth noting that they do not attempt to remove all AGN from their sample. 
Using the evolving luminosity function from \cite{Novak_2017} we expect to find $\sim 20$ SFGs in the redshift range $4.85<z<5.55$ and at $L_{\rm 1.4 \, GHz} > 10^{24}$\,W\,Hz$^{-1}$. 
Similarly, we can use the recent measurement of evolving radio luminosity function from \cite{Thykkathu2026}, who modelled the total radio luminosity function, including contributions from both the AGN and SFG populations, via a statistical approach similar to \cite{McAlpine2013}. 
From this we can estimate the number of SFGs and AGN in the area covered by the COSMOS-3D data. 
Using this, we find that the expected number of SFGs is $\sim 12$ and the number of AGN is $\sim 2$. 
There are significant uncertainties on these values, due to the fact that most optically-invisible sources are either omitted \citep{Novak_2017} in measuring the luminosity function, or have to be accounted for statistically \citep{Thykkathu2026}, and these are likely to be at the higher redshifts \citep[$z>3$, e.g.][]{Gentile2025}. 
However, they provide a reasonable estimate of the likely relative numbers of SFGs and AGN. 
Thus, we would expect to detect far more SFGs than AGN in these data, and this is consistent with the SFRs determined from both the radio continuum luminosity and other measurements of the SFRs.

\section{Conclusions}
\label{sec: conclusions}

In this work we present three high-redshift radio sources spectroscopically confirmed at $z=4.9$--$5.6$ detected in MIGHTEE radio continuum imaging at 1.3 GHz \citep{MIGHTEE_DR1}. 
These sources were identified by crossmatching the MIGHTEE DR1 continuum data to ground-based LBG catalogues at $z>4$ \citep{Adams23} and the COSMOS2025 catalogue \citep{COSMOS2025}.
These sources were identified as high-redshift candidates through SED fitting of \textit{Hubble} \textit{F814W}, \jwst NIRCam+MIRI, and ground-based UltraVISTA+Subaru HSC photometry taken from COSMOS2025.
The photometric redshifts of these sources were confirmed through a detection of the $\ha$ emission line in \jwst COSMOS-3D wide field slitless spectroscopy.
The radio luminosities of these sources are at least two orders of magnitude below previously identified HzRSs, with $L_{\rm 1.3 \, GHz}\approx2$--$5\times10^{24} \, \rm W \, Hz^{-1}$.
They have steep spectral indices, $\alpha^{\rm 3 \, GHz}_{\rm 1.3 \, GHz}\approx0.96$--$1.68$, determined using VLA 3 GHz data \citep{Smolcic2017}.

These HzRSs have S\'ersic indices and effective radii consistent with the size-mass relations of star-forming galaxies at $z\simeq5$ \citep{Varadaraj24, Allen25}.
None of the sources have a dominant point-source component (as expected for LRDs), indicating that AGN do not dominate their rest-frame UV and optical flux.
Two of the sources, \sfglong and \agnlong, show complex multi-component morphologies, which may suggest that they are undergoing mergers.
The third source, \starburstlong, shows a bulge+disk morphology where the bulge contributes no more than 20 per cent to the total flux.

We compute the SFRs based on SED fitting, $\ha$, and the 1.3 GHz luminosity, assuming that all of these tracers are powered by star formation.
The SFRs span $\approx100$--$1800\, \rm M_{\odot} \, \rm yr^{-1}$.
When averaging the SED-derived SFRs over 10 Myr and 100 Myr, we find that $\rm SFR^{SED}_{\rm 10 \, Myr}$ is larger than $\rm SFR^{SED}_{\rm 100 \, Myr}$, caused by a recent starburst in the SFH.
The SFRs from $\ha$ and the 1.3 GHz luminosity are consistent with, or lie between, $\rm SFR^{SED}_{\rm 10 \, Myr}$ and $\rm SFR^{SED}_{\rm 100 \, Myr}$ and lie either on or 0.5-1.0 dex above the star-forming main sequence at $z=4$--$6$.

The expected number of radio-loud SFGs and AGN based on the evolving radio luminosity function aligns with the three spectroscopically confirmed sources we find.
Additionally, we detect $\ha$ emission from three out of 18 HzRS photometric candidates which overlap with the COSMOS-3D footprint. 
This detection rate is broadly consistent with the timescales probed by $\ha$ ($\sim 10$ Myr) and the 1.3 GHz luminosity ($\sim 100$ Myr).

We interpret these HzRSs as radio sources which are undergoing starbursts, allowing for the detection of the $\ha$ emission line.
The two HzRSs with complex morphologies may represent merger-triggered starbursts.

The observations presented in this work represent the first detections of star formation-powered selected via their radio emission at $z>4.5$, demonstrating the power of combining ultra-deep radio continuum surveys in well-studied extragalactic fields with deep ground- and space-based imaging and spectroscopic datasets. Constraining the star formation, independent of dust, in some of the earliest galaxies that formed in the Universe is paramount to studying the assembly of the first structures after the Big Bang, and these observations lay the foundations for future work towards this goal.

\section*{Acknowledgements}

RGV, SF and MJJ acknowledge support from a UKRI Frontiers Research Grant [EP/X026639/1]. AS acknowledges
funding from the ``FirstGalaxies'' Advanced Grant from the European Research Council (ERC) under the European Union’s Horizon 2020 research and innovation programme (Grant agreement No. 789056).
IHW, CLH and MJJ acknowledge support from the Oxford Hintze Centre for Astrophysical Surveys which is funded through generous support from the Hintze Family Charitable Foundation. 
RAM acknowledges support from the Swiss National Science Foundation (SNSF) through project grant 200020\_207349.
KK acknowledges support from VILLUM FONDEN (71574) and the Danish National Research Foundation under grant no. 140.
JBC  acknowledges funding from the JWST Arizona/Steward Postdoc in Early galaxies and Reionization (JASPER) Scholar contract at the University of Arizona.
ZJL acknowledge the National Key R\&D Program of China (MOST) with grant No.\ 2022YFA1605300, and the National Natural Science Foundation of China (NSFC, grant No.\ 12273051).

\section*{Data Availability}
The MIGHTEE continuum DR1 data used in this work are released with \citet{MIGHTEE_DR1}, the MIGHTEE continuum Early Science cross-matched catalogue is realised with \citet{Whittam24}, and the MIGHTEE continuum Early Science host galaxy classifications are released with \citet{Whittam22}. The ground-based galaxy catalogues are released with \citet{Adams23}.
The \jwst COSMOS-Web data used in this work are released with \citet{COSMOS2025}.



\bibliographystyle{mnras}
\bibliography{bibliography} 




\appendix


\bsp	
\label{lastpage}
\end{document}